\documentclass{elsart5p}
\usepackage[authoryear]{natbib}
\usepackage{graphicx}
\usepackage{amssymb}
\usepackage{amsmath}
\usepackage{url}

\usepackage[dvipsnames,usenames]{color}
\begin{document}
\begin{frontmatter}
\title{Species abundances and lifetimes: from neutral to
  niche-stabilized communities}

\author[UPC]{Simone Pigolotti}
\and
\author[ISC]{Massimo Cencini}
\address[UPC]{Dept. de Fisica i Eng. Nuclear, Universitat
  Politecnica de Catalunya Edif. GAIA, Rambla Sant Nebridi 22, 08222
  Terrassa, Barcelona, Spain}
\address[ISC]{Istituto dei Sistemi Complessi (ISC), CNR, Via dei
Taurini, 19 00185 Rome, Italy}

\begin{abstract}
  We study a stochastic community model able to interpolate from a
  neutral regime to a niche partitioned regime upon varying a single
  parameter tuning the intensity of niche stabilization, namely the
  difference between intraspecific and interspecific competition. By
  means of a self-consistent approach, we obtain an analytical
  expression for the species abundance distribution, in excellent
  agreement with stochastic simulations of the model. In the neutral
  limit, the Fisher log-series is recovered, while upon increasing the
  stabilization strength the species abundance distribution develops a
  maximum for species at intermediate abundances, corresponding to the
  emergence of a carrying capacity.  Numerical studies of species
  extinction-time distribution show that niche-stabilization strongly
  affects also the dynamical properties of the system by increasing
  the average species lifetimes, while suppressing their
  fluctuations. The results are discussed in view of the niche-neutral
  debate and of their potential relevance to field data.
\end{abstract}


\end{frontmatter}

\section{Introduction}

Understanding the forces shaping ecological communities and promoting
species coexistence is a key longstanding issue in theoretical
ecology. Niche theory \citep{Macarthur1967a,Chesson2000,Chase2003} and
neutral theory \citep{Caswell1976,Bell2001,Hubbell2001} propose two
alternative mechanisms for explaining long-term coexistence of species
communities.

Many observational studies have highlighted the importance of
specificities in resource exploitation and thus of niche
differentiation for long-term coexistence \citep[see ][for two recent
reviews]{Wright2002,Silvertown2004}.  From a theoretical point of
view, niche partitioning enhances competition with conspecific individuals. As a
consequence, species limit their own populations more than those of
other species, providing a stabilization mechanism which promotes
coexistence \citep{Chesson2000}.  A simple mathematical formalization
of this idea is provided by the Lotka-Volterra competition model
\citep{Macarthur1967a}, where coexistence emerges whenever
intraspecific interactions overtake interspecific ones.

Conversely, ecological neutral theory
\citep{Hubbell2001} completely disregards species differences and
assumes functional equivalence at the individual level.  Within this
theoretical approach, stochasticity is the leading ecological force,
and species coexistence is the result of a dynamical balance between
immigration/speciation processes and extinction.  
The drastic assumptions of neutral theory favour
  mathematical tractability leading to analytical predictions for
  ecological patterns such as species abundance distribution
  \citep[see, e.g.,][and references therein]{Rosindell2011}.
Such predictions, depending on few parameters, fit surprisingly well
field data from tropical forests \citep[see,
  e.g.,][]{Hubbell2001,Volkov2003,Volkov2005}.

Due to its strong departure from traditional theoretical approaches,
the neutral theory elicited a heated debate in community ecology about
its validity and interpretation. A complication is that,
in many cases, niche-based and neutral models yield similar fits of
biodiversity patterns \citep[see,
e.g., ][]{Chave2002,McGill2003,Mouquet2003,Tilman2004,McGill2006}.

Here, we do not enter this debate, but embrace the view that ``niche
and neutral models are in reality two ends of a continuum with the
truth most likely in the middle'' \citep{Chase2003}.  Indeed, the
ecological forces underlying niche and neutral models are not mutually
exclusive.  Following a similar view, different authors have proposed
models synthesizing niche-based and neutral mechanisms
\citep{Tilman2004,Leibold2006,Gravel2006,Adler2007,Kadmon2007,Pigolotti2010,Haegeman2011,Noble2011a,Noble2011b}.
A desired property of these models is the possibility to recover pure
neutrality as a limiting case \citep{Alonso2006}, allowing for
studying the transition from niche-dominated to neutral regime when
one (or more) ecological parameter is varied. A related issue concerns
the robustness of the neutral theory.  In particular, a general
understanding of whether small non-neutral effects can undermine
neutral predictions is still lacking. Finally, niche-neutral models
could help our understanding of the dynamical features and long time
behavior of ecosystems. Indeed, while neutral theory in principle
allows for extrapolating ecosystem behavior at long time-scales
\citep{Pigolotti2005,Azaele2006,Bertuzzo2011}, an open issue is that
predicted species lifetimes tend to be too short compared to fossil
records estimates \citep[see,
  e.g.][]{Lande2003,Ricklefs2003,Nee2005,Ricklefs2006,Allen2007}

A difficulty with the above program is that usually such mixed models
are very hard to analyze mathematically \citep[see,
  e.g.,][]{Tilman2004}. In this respect, surely, one of the reasons of
success of the neutral theory has been its mathematical tractability,
i.e. the fact that closed expression for distribution such as species
abundance distributions (SAD) can be obtained and easily compared with
data \citep{Volkov2003,Volkov2005}. This is usually not possible in
models where the neutral hypothesis is broken. Moreover, breaking the
neutral hypothesis often leads to a proliferation of parameters,
making general comparisons problematic.

As discussed by \citet{Adler2007}, neutrality can be violated both at
the level of fitness inequalities among the species and/or by the
presence of stabilizing processes causing species to limit themselves
more than they limit others. These are the two main mechanisms for
coexistence identified by \citet{Chesson2000}.  As for the former kind
of violation, \citet{Zhang1997,Zhou2008} have shown that even small
differences in fitness in an otherwise neutral model cause strong
reduction of biodiversity. This result is not unexpected as fitness
differences, in the absence of stabilizing effects, clearly lead to
competitive exclusion. As for the second kind of violation of
neutrality, quite interesting is the approach followed by
\citet{Haegeman2011} who added demographic stochasticity and
immigration to classical competitive Lotka-Volterra dynamics with
symmetric interactions.  A more general model, introduced in
\citep{Noble2011a,Noble2011b,Noble2011c}, incorporates both fitness
inequalities and stabilizing mechanisms. In a nutshell, Noble et
al. added a frequency-dependent birth rate to the Moran model in a
such a way that it reduces to Hubbell's neutral model in an
appropriate limit and recovers standard Lotka-Volterra phenomenology
in the deterministic limit. They mostly explored ecosystems with few
species, aiming at understanding niche stabilization effects at
intermediate time scales.

In this paper we investigate a variant of the model introduced by
\cite{Noble2011b} \citep[see also][]{Noble2011c}, with stochastic
death and reproduction events where fitness inequalities are
disregarded while stabilizing mechanisms are retained via symmetric
interactions, similarly to \citep{Haegeman2011}.  Unlike
\citet{Noble2011b} and \citet{Haegeman2011} we focus on large
metacommunities with possibly many species subject to extinction and
speciation, allowing us to probe the effect of niche-stabilization
both on static patterns, such as the species abundance distribution,
and dynamical ones, such as the species extinction time
distribution. A similar study aimed at testing the role of
niche-stabilization was undertaken in \citet{Adler2010}, where a
suitable individual based model was tuned by fitting the parameters of
the dataset of a sagebrush steppe community. There, only four species
were present and speciation was not considered. It is however
remarkable that the sagebrush community there investigated was not too
far from our model assumptions, namely equal fitness and symmetric
competition.  Furthermore, other field-data studies
\citep[see. e.g.,][for a recent study in this direction]{Volkov2009}
support the idea that the most relevant deviation from neutrality
consists in differences between intraspecific and interspecific
competition. Such differences are usually explained in terms of
species-specific resource consumption or more sophisticated mechanisms
like Janzen-Connell effects \citep{Janzen1970,Connell1971}. Therefore,
considering, as done here, equal fitness and intraspecific interaction
larger than interspecific one is not too restrictive. The symmetric
assumption may instead be stronger but has, at least, some relevance
to field data.

One of our main result here is to derive a closed analytical
expression for the SAD for the model we study.  As this model is more
complex than its neutral counterpart, our calculations rely on some
assumptions and approximations. However, the obtained analytical
species abundance curves are nearly indistinguishable from
individual-based simulations of the model in a wide parameter range,
supporting the validity of our approach. In the neutral limit, the
classical Fisher log-series is recovered
\citep{Fisher1943,Hubbell2001}. Departing from the neutral limit, the
effect of niche stabilization is essentially reabsorbed into a single
self-consistently derived parameter, which also modifies Hubbell's
fundamental biodiversity parameter. As discussed before, most existing
datasets satisfying our hypothesis are either made up of few species,
or simply well fitted by neutral theory. Nevertheless, the
availability of an explicit closed formula can be useful for
fitting SAD in future field studies of diverse communities, where one
believes that the impact of niche stabilization cannot be neglected.

We also quantify the effect of stabilization on species extinction
times. Here, the extinction time --- sometimes also dubbed species
persistence time \citep{Bertuzzo2011} --- is defined as the time from
appearence of the first individual entering the system via speciation
to the extinction of the last individual of such species.  In the
neutral case, the probability distribution of such extinction time can
be analytically computed \citep{Pigolotti2005}. In the presence of
niche stabilization, we could not obtain an analogous closed
expression, so that we resorted to stochastic simulations at varying
the niche-stabilization parameter.

Model details are summarized in Sec.\ref{sec:model}.
Section~\ref{sec:results} presents the main results: the analytical
derivation of the SAD with numerical validation, and a numerical study
of the species extinction times distribution as a function of the
deviation from neutrality. Section~\ref{sec:discussions} is devoted to
discussing the results. The Appendix contains some technical details
of the self-consistent derivation presented in Sec.~\ref{sec:results}.

\section{Model\label{sec:model}}

For the sake of simplicity, we start by introducing the neutral
version of the model, which is the standard Moran model for a
community with a fixed number $J$ of individuals. At each timestep
an individual, randomly chosen among the $J$ composing the community,
is killed. With probability $\nu$, it is replaced with an individual
belonging to a new species, not present in the system
(speciation). With probability $(1-\nu)$, it is replaced with a copy
of one of the $J-1$ individuals already present in the community
(reproduction). Denoting with $n_i$ the number of individuals of
species $i=1,\ldots,S$ ($S$ being the number of species currently in
the system) present at time $t$, the death of an individual of species
$j$ followed by the birth of one of species $i$ happens at rate
\begin{equation}
(1-\nu) \frac{n_j}{J} \frac{n_i}{J-1}\,,
\label{eq:neutral}
\end{equation}
with $n_i,n_j\geq 1$.

We now move to the general non-neutral case.  Similar to
\citet{Noble2011b}, non-neutrality is introduced by biasing
reproduction. We introduce the frequency-dependent weights
\begin{equation}\label{m2}
w_i =r_ie^{- \sum_{j=1}^S a_{ij}n_j}\,,
\end{equation}
and substitute the term $n_i/(J-1)$ of Eq.~(\ref{eq:neutral}) with
\begin{equation}
\frac{w_i\,n_i}{\sum_{k=1}^{S}  w_k n_k}\,,
\label{eq:noble}
\end{equation}
where $r_i$ models the intrinsic fitness advantage of species $i$ and
$a_{ij}$ the competition among the species present in the
community. With equal fitness $r_i=r$ and uniform competition
$a_{ij}=a_j$, Eq.~(\ref{eq:noble}) reduces to the neutral value
$n_i/(J-1)$  (by definition $\sum_{k=1}^{S} n_k=J-1$ as the
dead individual does not participate to reproduction). In a different
limit, one can draw a correspondence with the Lotka-Volterra
competitive model by expanding (\ref{eq:noble}) for small $a_{ij}$ and
neglecting fluctuations \citep{Noble2011b,Noble2011c}. 

Specifically, we consider a simplified variant of the model by
imposing equal fitness $r_i=r$ while retaining niche-stabilization
\citep{Adler2007} through a fully symmetric competition matrix
$a_{ij}=a\delta_{ij}+b(1-\delta_{ij})$, as in \citet{Haegeman2011}.
With these simplifications, Eq.~(\ref{eq:noble}) reduces to
\begin{eqnarray}\label{om_def_old}
\frac{n_ie^{-an_i-b\sum_{j\neq i}n_j}} {\sum_{k=1}^S n_k e^{-an_k-b\sum_{j\neq k}n_j}}  
=\frac{n_ie^{-(a-b)n_i}}{\sum_{k=1}^S n_k e^{-(a-b)n_k}}\,,
\end{eqnarray}
where we used $\sum_{j\neq i} n_j=(J-1)-n_i$.  Parameters $a$ and $b$
tune the weights of intraspecific and interspecific competition,
respectively. Finally, by defining
$c=a-b$, the frequency-dependent rates can be written as
\begin{equation}\label{om_def}
\omega^{(i)}(n_i)=\frac{n_ie^{-cn_i}}{\sum_{k=1}^S n_k e^{-cn_k}}\,,
\end{equation}
where $c$  quantifies the deviation from the
neutral case, $c=0$. We focus on the case niche stabilization $c>0$,
meaning intraspecific competition being stronger than interspecific
one (i.e. $a>b$), as observed in real data \citep{Volkov2009}.

\section{Results\label{sec:results}}

In this section we present the main results. First, we show that the
species abundance distribution (SAD) of the model can be analytically
computed by means of an approach similar to that used in
\citet{Volkov2003} for the neutral model, complemented with a
self-consistent ansatz for the treatment of non-neutral interactions.
Second, we study the dynamical properties of the model by numerically
investigating the species extinction times statistics.
 
\subsection{\textit{Species abundance distribution}} 

The empirical Species Abundance Distribution (SAD) $\phi(k)$ counts,
among all species $S$ present in a specific sample ecosystem, those
having $k$ individuals.  In formulas, this can be expressed as
$\phi(k) = \sum_{i=1}^{S} \delta_{k,n_i}$, with the Kroneker symbol
$\delta_{km}=\!\!1$ if $k=m$ and zero otherwise.  We are interested in
the average SAD, $\Phi(k)=\langle \phi(k)\rangle$, where the brackets
$\langle [\ldots] \rangle$ denote an average over time or,
equivalently, over different realizations of the metacommunity
dynamics.

In the following, we assume that species can be treated as
independent,\footnote{It is important to stress that the assumption of
  independence does not necessarily amount to neglecting
  interactions. Indeed, this approach is commonly (and successfully)
  adopted in physics and chemistry to tackle complex many-body
  problems, such as in the so-called Hartree-Fock method or other
  mean-field approximations: instead of trying to include interactions
  exactly, one formally describes each particle as independent, but
  feeling an averaged effect of the interactions with the other
  particles. This will be the idea underlying the crucial step of our
  derivation. } so that the average can be performed over
single-species abundance distributions.
Furthermore, we assume that such probabilities do not
depend on the species label as a consequence of species
symmetry. Then, upon defining the probability $P_{k}\equiv \langle
\delta_{k,n_i} \rangle$ of a species having $k$ individuals, the mean
SAD can be expressed as
\begin{equation}
\Phi(k)= \langle \phi(k)\rangle = \left\langle \sum_{i=1}^{S} \delta_{k,n_i}
\right\rangle  = \langle S\rangle P_k\,,
\label{def:sad-mean}
\end{equation}
where $\langle S\rangle$ is the average number of species
present in the system. The probabilities $P_k$ are normalized to one,
$\sum_{k=1}^{\infty} P_k=1$, while the SAD sums up to the average
number of species, $\sum_{k=1}^{\infty} \Phi(k)=\langle S\rangle$. We also
remark that in the original model the total number of individuals is
fixed and equal to $J$.  This constraint is lost due to the
independent-species assumption, but we will reintroduce it at the end
of the calculation by imposing $\sum_{k=1}^{\infty} k \,\Phi(k)=J$.

\begin{figure*}[th!]
\centering
\includegraphics[width=0.9\textwidth]{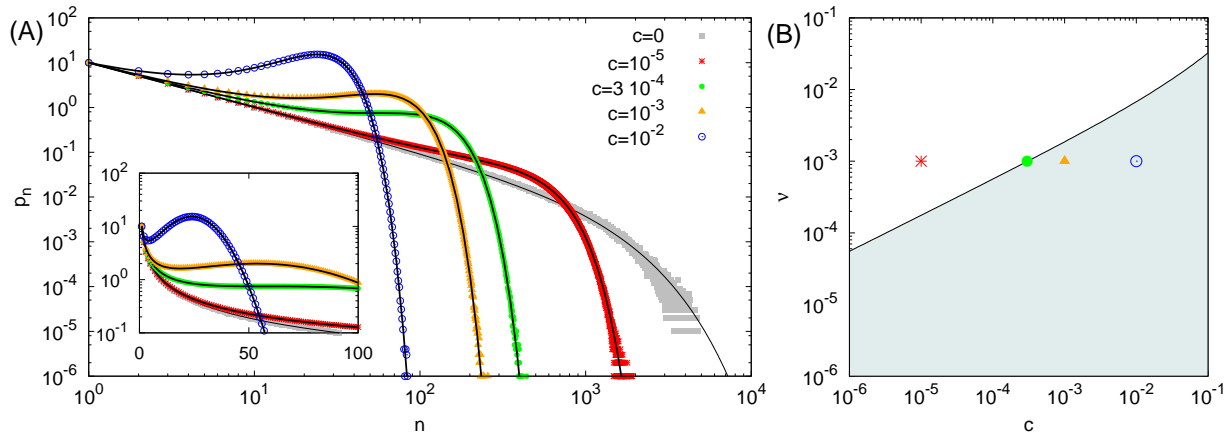}
\caption{(A) Species abundance distribution at varying the stabilization
  parameter $c$ as in the legend. Inset: zoom of the same curves in log-lin
  scale to highlight the emergence of a mode. Symbols refer to
  simulations and solid curves to the prediction (\ref{solution}). (B)
  The $(c,\nu)$ parameter plane. The grey region corresponds to
  parameter values for which the condition (\ref{eq:condition}) is
  fulfilled, i.e. where a peak in the SAD is present. The symbols
  correspond to parameters as in the simulations shown in (A) but for
  the neutral value $c=0$ here not represented. \label{fig1}}
\end{figure*}

Typical SAD $\Phi(n)$ generated by the considered model are shown in
Fig.~\ref{fig1}A for a system of $J=10^4$ individuals with speciation
rate $\nu=10^{-3}$ and different values of $c$. Increasing the
stabilization parameter $c$, we observe a transition from a
neutral-like SAD, given by the Fisher-log series \citep{Fisher1943},
to curves characterized by a maximum at intermediate abundances.  The
remainder of this section is dedicated to the derivation of an
analytical expression for $\Phi(n)$.

As it will be used later, we introduce the Laplace transform of the
SAD:
\begin{equation}
\widetilde{\Phi}(z)= \sum_{k=1}^{\infty} \Phi(k) e^{-zk}\,,
\label{eq:generating}
\end{equation}
in terms of which the normalization to a prescribed number of individuals can
be rewritten as
\begin{equation}
\widetilde{\Phi}^{\prime}(0)= \sum_{k=1}^{\infty} k\Phi(k)=J\,,
\label{eq:generatingJ}
\end{equation}
the prime denoting the derivative with respect to $z$.

In order to find an explicit analytical expression for $\Phi(k)$ using
Eq.~(\ref{def:sad-mean}) we need the equilibrium probabilities
$P_k$. To derive an expression for such probabilities we now consider
a slightly simplified version of the model introduced in
Sect.~\ref{sec:model}. We focus on a single species, labeled $i$, and
write the master equation, 
\begin{equation}
\frac{d}{dt} p^{(i)}_n(t)=b_{n-1}^{(i)}p^{(i)}_{n-1} +d^{(i)}_{n+1}p^{(i)}_{n+1}-(b^{(i)}_n+d^{(i)}_n)p^{(i)}_n\,,
\label{eq:master-equation}
\end{equation}
ruling the evolution of the probability $p^{(i)}_n(t)$ of species $i$
to have $n_i=n$ individuals at time $t$. The death and birth
rates, $d_n^{(i)}$ and $b^{(i)}_n$ respectively, can
be expressed as follows. The former is completely neutral and
determined by the frequency of species $i$
\begin{equation}
d_n^{(i)}=\frac{n_i}{J}\,.
\label{eq:dn}
\end{equation} 
The latter, instead, explicitly accounts for competition with the
other species 
\begin{equation}
b_n^{(i)}=(1-\nu)\omega^{(i)}(n_i)=(1-\nu)\frac{n_ie^{-cn_i}}{\sum_{k=1}^S n_k e^{-cn_k}}\,.\label{eq:bn}
\end{equation}
As species are interacting, this probability depends parametrically on
the populations of all the other species. The main difference between
this simplified dynamics and the original model encoded in
Eqs. (\ref{eq:neutral}), (\ref{m2}) and (\ref{eq:noble}) is that we
are not taking into account that birth and death events are
simultaneous at the community level. Consequently, the constraint on
the total number of individuals is lost. As already discussed above,
this constraint will be reintroduced with a suitable normalization of
the SAD. In Eq.~(\ref{eq:bn}), the birth probability $\omega^{(i)}(n)$
is decreased by the factor $(1-\nu)$ to account for speciation, which
from the point of view of the focal species corresponds to the
possibility to give birth to an individual of a new species. Following
\citet{Volkov2003} we set $b^{(i)}_0=\nu$. Notice that imposing
$b^{(i)}_0>0$ is necessary to achieve a proper stationary solution
(see \citet{Azaele2006b} for mathematical details), while its actual
value is irrelevant as it can be reabsorbed into the normalization
condition.

If the populations of the other species, $n_j$ for
  $j\neq i$, were constant, the equilibrium distribution
$p^{eq(i)}_k$ for species $i$ would simply be
obtained by imposing detailed balance
\begin{equation}
\label{detbal}
p^{eq(i)}_k = \mathcal{N} \prod_{j=1}^k \frac{b^{(i)}_{j-1}}{d^{(i)}_j}
=  \mathcal{N}\frac{\nu}{1-\nu}\, \dfrac{\displaystyle{J^k(1-\nu)^ke^{-\frac{c}{2}k(k-1)}}}
       {\displaystyle{k\prod_{j=1}^k \, \sum_{l=1;n_i=j}^S n_le^{-cn_l}}}\,
\end{equation}
where the constant $\mathcal{N}$ has to be fixed by the normalization
condition, $\sum_{k=1}^{\infty} p^{eq(i)}_k=1$.  We stress that,
strictly speaking, (\ref{detbal}) represents just a formal expression
for the equilibrium probability of species $i$, $p^{eq(i)}_k$, valid
for given values of the populations of the other species $n_l$ (with
$l\neq i$).

Now we make the ansatz --- whose validity will be justified a
posteriori by comparing the result of the calculation with numerical
data --- that Eq.~(\ref{detbal}) can be used to express the
equilibrium probabilities, i.e. $p_k^{eq(i)}=P^{(i)}_k$, of the
considered model by replacing the denominator with the a suitable
average over the probabilities to be determined.  Doing so we will
obtain a self-consistent expression for the SAD.

We exploit again species symmetry by dropping any
  explicit dependence on the species label $i$. Then, consistently,
we relax the constraint $n_i=j$ in the sum appearing in the
denominator of (\ref{detbal}) and rewrite it as an
  average expressed in terms of the (yet unknown) SAD
\begin{eqnarray}
&&\left\langle \sum_{j=1}^{S} n_j e^{-cn_j} \right\rangle = \left\langle \sum_{j=1}^{S} \sum_{k=1}^\infty
\delta_{k,n_j} n_j e^{-cn_j}\right\rangle \nonumber \\&&= \sum_{k=1}^\infty k e^{-ck} \left\langle\sum_{j=1}^{S} 
\delta_{k,n_j}  \right\rangle  =\sum_{k=1}^{\infty} k \Phi(k) e^{-ck}=-\widetilde{\Phi}^\prime(c)\,,
\label{eq:approx1}
\end{eqnarray}
where we used (\ref{def:sad-mean}) in the third
  equality and (\ref{eq:generating}) in the last one.  Now, using
Eqs.~(\ref{def:sad-mean}), (\ref{detbal}), and (\ref{eq:approx1}), we can
write
\begin{equation}
\Phi(k) = \theta \dfrac{J^k(1-\nu)^ke^{-\frac{c}{2}k(k-1)}}
       {k \left(\sum_{j=1}^{\infty} j \Phi(k) e^{-cj} \right)^k}
\label{eq:sad-teo}
\end{equation}
where the constant $\theta$ will be fixed by imposing
$\sum_{k=1}^{\infty}k \Phi(k)=J$, similarly to Hubbell's biodiversity
number in the neutral case \citep{Volkov2003}.  Notice that
(\ref{eq:sad-teo}) is a self-consistent equation
  expressing $\Phi$ as a function of $\Phi$ itself. Finally, using
(\ref{eq:generatingJ}) to re-express $J$, the last equality in
(\ref{eq:approx1}), and introducing the constant
\begin{equation}
g={\widetilde{\Phi}^\prime(0)}/{\widetilde{\Phi}^\prime(c)}
\label{def:g}
\end{equation}
we obtain our central result
\begin{equation}
\Phi(k) = \theta \dfrac{[(1-\nu)g]^ke^{-\frac{c}{2}k(k-1)}}{k} \,.
\label{solution}
\end{equation}
The above equation provides an analytical prediction for the SAD in
terms of the unknown constant $g$ which needs to be determined
self-consistently. From the definition (\ref{def:g}) we obtain the
self-consistent condition
\begin{equation}\label{gdef1}
g=g(c,\nu)=\frac{\widetilde{\Phi}'(0)}{\widetilde{\Phi}'(c)}=
\frac{\sum_{k=1}^\infty (1-\nu)^kg^k e^{-\frac{c}{2}k(k-1)}}
{\sum_{k=1}^\infty (1-\nu)^kg^k e^{-\frac{c}{2}k(k+1)}}\,.
\end{equation}
As shown in the Appendix, the above equation can be approximated in
the continuum limit and $g$ can be obtained as the solution of the
transcendental equation
\begin{eqnarray}
\frac{1-\nu}{\nu}&=&\sqrt{\frac{\pi}{2c}}\mathrm{erfc}\left(\frac{3c-2\ln(g(1-\nu))}{2\sqrt{2c}}\right)e^{{\frac{[c-2\ln((1-\nu)g)]^2}{8c}}}\,. \label{g_sem}
\end{eqnarray}
Further, imposing the normalization $\sum_{k=1}^{\infty}k \Phi(k)=-\Phi^{\prime}(0)=J$ leads to (see Appendix)
\begin{equation}\label{norm}
\theta=\frac{J\nu}{g(1-\nu)}\,.
\end{equation}

In the neutral limit, $c=0$, we have $g=1$ so that $\theta$ is equal
to the neutral biodiversity number $J\nu/(1-\nu)$ and the SAD
(\ref{solution}) reduces to the Fisher log-series
\citep{Hubbell2001,Volkov2003}. Conversely, in the general case
$c\neq0$, $g>1$ so that the shape of the SAD will vary and the
biodiversity number is reduced with respect to the neutral case.
However, as shown below, such reduction does not imply lower
diversity, as the average number of species always increases with
niche-stabilization.

In Figure~\ref{fig1}A, SADs from the model at different values of
$\nu$ and $c$ (as indicated by symbols in the $c,\nu$ plane shown in
Fig.~\ref{fig1}B), are compared with the analytical prediction of
Eq. (\ref{solution}), represented in solid lines.  In all cases,
numerical data (symbols) are in excellent agreement with the theory,
validating the assumptions underlying the self-consistent approach.

For small values of $c$, the SAD is monotonically decreasing with $n$,
similarly to the (neutral) Fisher log-series but for a faster (Gaussian)
fall-off for large populations. Interestingly, for larger values of
$c$, SAD curves develop a maximum at a finite number of individuals,
as emphasized in the inset of Fig.~\ref{fig1}A. A study of the
function (\ref{solution}) reveals that the condition for developing a
maximum at $n>1$ is
\begin{equation}
\frac{c}{2}+\ln(g(1-\nu))-2\sqrt{c}>0\,,
\label{eq:condition}
\end{equation}
corresponding to the shaded area in Fig.~\ref{fig1}B.

\begin{figure*}[t!]
\centering
\includegraphics[width=0.9\textwidth]{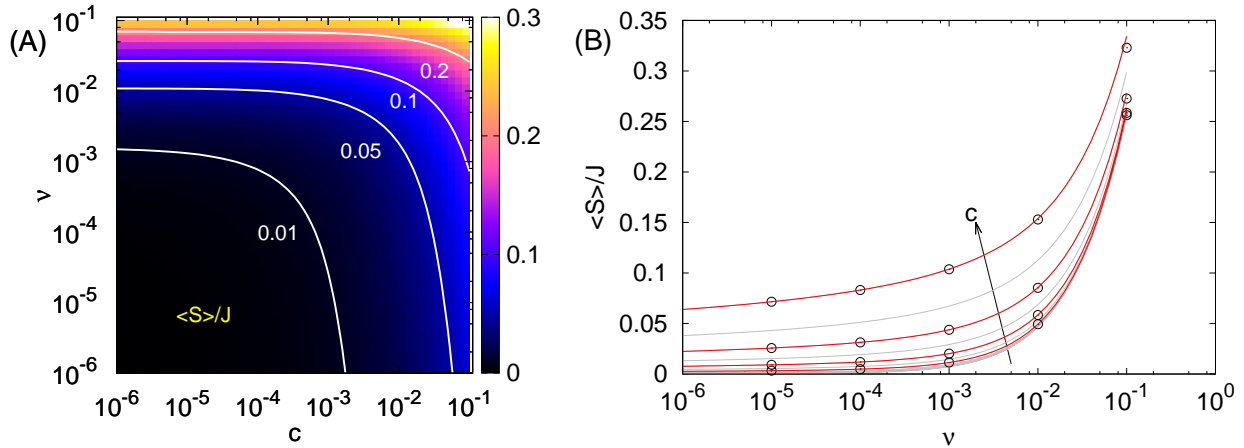}
\caption{(A) Average number of species normalized by the community
  size, $\langle S\rangle/J$, in the $(c,\nu)$ parameter space, as
  predicted by summing the terms of Eq.~(\ref{solution}). (B) Same
  quantity as a function of $\nu$ for different values of $c$, 
  increasing in the arrow direction. Symbols are results from
  simulations, in excellent agreement with the prediction in red
  lines. Grey lines correspond to values of $c$ different from those
  of the simulations. \label{fig2}}
\end{figure*}

We now discuss how the average number of species
$\langle S\rangle=\sum_{k}\Phi(k)$ changes upon
  varying the stabilization parameter. As the number of species is
proportional to the number of individuals $J$, in Fig.~\ref{fig2}A we
plot the fraction $\langle S\rangle /J$ as a function of the
speciation rate $\nu$ and the intensity of stabilization $c$.  For any
fixed value of the speciation rate $\nu$, the average number of
species grows {upon increasing $c$, meaning that the
system is able to sustain a larger diversity with respect to the
neutral case thanks to the stabilizing effect of niches. In
Fig.~\ref{fig2}B, we compare the prediction obtained by summing the
terms of Eq.~(\ref{solution}) with numerical simulation of the model,
again showing an excellent agreement between theoretical prediction
and simulation of the stochastic model.

\subsection{\textit{Species extinction times}} \label{turn_sec}

We now focus on dynamical, rather than static, patterns which provide
useful information on the differences between neutral and non-neutral
theories \citep{Pigolotti2005,Azaele2006,Allen2007,Bertuzzo2011}. In
particular, it is interesting to compare features of the dynamical
balance between appearance and extinction of species as a function of
the stabilization parameter $c$.  As in the model species originate
from a single individual, we are interested in the statistics of the
time it takes for a new species (introduced via a single individual)
to become extinct. The results are presented in unit of generations,
meaning that $t=1$ corresponds to $J$ iteration steps of the model of
Sec.~\ref{sec:model}.

In the neutral case $c=0$, by applying ideas from the theory of
branching processes \citep{Harris1989}, it is possible
to derive an analytical expression for the probability of a species to
survive a time $t$ (from the time of its introduction in the system),
which is given by \citep{Pigolotti2005}
\begin{equation}\label{extneut}
p_e(t)=\left(\frac{\nu}{e^{\nu t}-1+\nu}\right)^2e^{\nu t}.
\end{equation}
For $\nu\ll 1$, such distribution displays a power-law tail $t^{-2}$,
followed by an exponential cutoff set by $\nu$.  Notice that the
power-law $t^{-2}$ can be modified by dispersal properties different
from the global one here considered \citep{Bertuzzo2011}.  For $c>0$,
no closed expression for the extinction-time distribution is
available. Indeed, also a simple quadratic nonlinearity in the birth
(or death) rates for a single species makes the problem analytically
intractable, but for clever approximations as discussed in a series of
works \citep[see,
  e.g.][]{Norden2001,Newman2004,Doering2008,Parsons2008}. For this
reason, we limit ourselves to a numerical study.  In Figure
\ref{fig-pdf}, we show the extinction-time distribution, $p_e(t)$, for
different values of $c$ fixing $\nu$. Increasing $c$ has a non-trivial
effect on the shape of the extinction-time distribution.

Qualitatively, one can identify three regimes. At short times, the
distribution is only weakly dependent on $c$, i.e. by the presence of
stabilization, and essentially reproduces the neutral result. At
intermediate times, the extinction-time probability increases with
$c$. Finally, at large extinction times it decreases
  with $c$ by developing an exponential cutoff, $p_e(t) \sim
  \exp(-\alpha t)$, much steeper than predicted by formula
  (\ref{extneut}), i.e. with $\alpha(c)>\nu$.  The exponential rate of
  decrease, $\alpha$, as a function of the stabilization parameter $c$
  (shown in the inset of Fig.~\ref{fig-pdf}) suggests, at least for
  not too large $c$, a logaritmic dependence of extinction probability,
  namely $\alpha=a \ln c+b$, which would imply of the form $p_e(t) \sim c^{-a
    t} e^{-bt}$.

\begin{figure}[t!]
\centering
\includegraphics[width=0.45\textwidth]{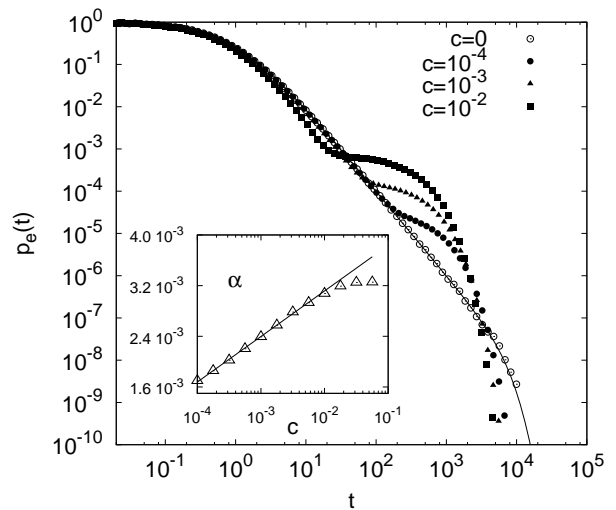}
\caption{Probability density function, $p_e(t)$, of the extinction
  times for $J=10^4$, $\nu=5\cdot 10^{-4}$ and different values of the
  stabilization parameter $c$, as in the legend. For $c=0$ the
  prediction of (\ref{extneut}) is perfectly verified. Inset: behavior
  of the exponential rate of decrease $\alpha$ of the pdf tail, where
  $p_e(t) \sim \exp(-\alpha t)$, as a function of $c$. Notice that the
  graph is in log-lin scale.  The solid line displays the fit
  $\alpha=a \ln(c)+b$ with $a= 0.000312$ and
  $b=0.00455$. \label{fig-pdf}}
\end{figure}

\begin{figure}[t!]
\centering
\includegraphics[width=0.45\textwidth]{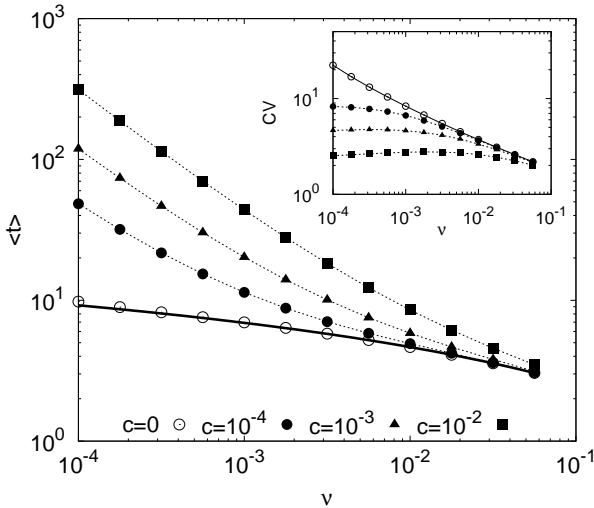}
\caption{Average extinction time as a function $\nu$ for different
  values of $c$, as in the legend. For $c=0$, the prediction of
  Eq. \ref{meanext} perfectly agrees with the data. Inset: coefficient
  of variation of the extinction time, defined as $CV=(\langle
  t^2\rangle-\langle t\rangle^2)^{1/2}/\langle t\rangle$, as a
  function of $\nu$ for the same values of $c$. \label{fig-mom}}
\end{figure}

As suggested by niche-stabilization arguments, the net effect of these
different regimes on the average extinction time is always positive:
the average lifetime of a species in our model is a growing function
of $c$ (see Fig.~\ref{fig-mom}), which can exceed up to 2 or 3 order of magnitude (depending on $\nu$) the  neutral value 
\begin{equation}\label{meanext}
\langle t\rangle = \frac{\ln(1/\nu)}{1-\nu}\,,
\end{equation}
derived from Eq.~(\ref{extneut}).  It is interesting to notice
that, while enhancing stabilization (larger $c$) increases
the average extinction time, it decreases its relative fluctuations as
confirmed by the behavior of the coefficient of variation,
$CV=\sqrt{\langle t^2\rangle - \langle t \rangle^2}/\langle t
\rangle$, shown in the inset of Fig.~\ref{fig-mom}. This means that,
while in the neutral case when $\nu\ll 1$ extinction times can vary
greatly around the average value, niche stabilization has also the
effect of making the average time a better prediction for the lifetime
of a randomly chosen species.

\section{Discussions\label{sec:discussions}}

We studied a community model incorporating neutral demographic
stochasticity with niche stabilization. The model belongs to a general
class of niche-neutral models discussed in the recent literature
\citep{Gravel2006,Adler2007} and, in particular, is built upon two
specific models originally proposed by \citet{Noble2011b,Haegeman2011}
to which we added speciation.  We assumed that violation of neutrality
is controlled by a single parameter, $c$, tuning the intensity of
niche stabilization. For the sake of simplicity, similarly to
\citet{Haegeman2011}, we assumed that, even in non-neutral cases,
species are symmetric and with no intrinsic fitness advantages, as
also observed in some field data \citep{Adler2010}. The model depends
also on rate $\nu$ of introduction of new species by speciation or
immigration.

We proposed a self-consistent calculation which allowed us to derive
an analytical expression for the species abundance distribution,
Eq.~(\ref{solution}).  The assumptions and approximations at the basis
of the self-consistent calculation seem to be well verified and the
analytical formula does not show significant deviations from
simulations of the individual based model.  While we focused here on
one specific source of non-neutrality, possibly other non-neutral
(symmetric) models may be approached using similar ideas.  Indeed,
species permutation symmetry is the crucial assumption to compute the
SADs. However, we remark that even in the presence of species
permutation symmetry, the self-consistent ansatz may not lead to the
correct result when this symmetry is spontaneously broken
\citep{Borile2012}.

Interestingly, in a large portion of the parameter space $(c,\nu)$,
the relative abundance distribution curves show a maximum for
intermediate abundance classes.  Specifically, such mode appears when
the stabilization parameter $c$ exceeds a critical value (predicted by
Eq.~(\ref{eq:condition})) depending on the rate $\nu$ of introduction
of new species. This feature of the distributions we obtain is
reminiscent of many instances of SAD curves measured on the field
\citep{Hubbell2001}, although one should keep in mind that when
plotting in log abundances, as in a Preston plot, the mode will emerge
at a quantitatively different value of $c$ than predicted from
Eq.~(\ref{eq:condition}).  We remind that spatially implicit
implementations of the neutral model, which effectively include a form
of dispersal limitation by coupling a community and a metacommunity,
provide very good fits of such field-measured SAD curves
\citep{Hubbell2001,Volkov2003}. Without dispersal limitation, good
fits have been obtained by combining neutral community models with
specific forms of density-dependent reproduction rates
\citep{Volkov2005}.

Within the model here investigated, the presence of a peak in SAD curves
corresponds to the natural emergence of an effective carrying capacity
for each niche, so that the population of each species fluctuates
around a well defined average value. It should be remarked, however,
that the tail of the SAD distribution, i.e. the probability of
encountering a very abundant species, falls off in a much sharper way
for large $n$ than in Hubbell's spatially-implicit neutral model, due
to the presence of a term proportional to $\exp(-cn^2)$ in
Eq.~(\ref{solution}).  Typical datasets  well fitted by the
neutral theory \citep{Hubbell2001,Volkov2003} support the presence of a fatter
log-normal-like tail. As a consequence, best fits of those datasets
with formula~(\ref{solution}) (not shown) are biased by such long
tails and are thus realized close or at the neutral limit (i.e. for
$c\rightarrow 0$), where the peak disappears.

The emergence of this Gaussian cutoff, conflicting with classic
datasets, is likely due to the fact that the niche-stabilization
mechanisms incorporated in our model are particularly strong in the
absence of dispersal limitation. It is indeed reasonable to expect
that the sharp fall-off should be a quite general feature for global
dispersal models in which an effective carrying capacity for any
single species emerges and should not depend too much on model
details.  We conjecture that including some form of dispersal
limitation with  competition acting on a finite range may lead to
fatter tails. Similar features were observed in \citet{Chave2002}
while comparing global and limited dispersal models with density
dependence or other stabilizing mechanisms, such as tradeoffs.  In
this perspective, it would be interesting to extend the model to
incorporate, even in a spatially implicit form, some degree of
dispersal limitation. Unfortunately, this extension makes the
analytical treatment much harder and it is left for future
investigations. Another way fatter tails may arise would be to break
the symmetric hypothesis and letting different species having
different carrying capacity.

Studying dynamic patterns of species lifetimes (Sect.~\ref{turn_sec})
revealed that the main effect of violating neutrality via
niche-stabilization is to suppress the large fluctuations of
lifetimes, typical of neutral dynamics \citep{Pigolotti2005}.  On the
one hand, stabilization prevents the possibility of a species to
achieve a very large population size and thus lifetimes much longer
than the average, which are exponentially suppressed.  Similar
exponential distributions of lifetimes have been observed in fossil
data, a fact often explained in terms of the Red Queen effect in a
changing environment \citep{VanValen1973}.  On the other hand, a
species is favored when rare making it less prone to extinction by
demographic stochasticity when its population is small \citep[see
  also][]{Adler2010}.  These effects are weighted by the stabilization
parameter, as a result the average species lifetime results increased
with respect to the neutral expectation (\ref{meanext}).

As observed by many authors \citep[see,
  e.g.][]{Lande2003,Ricklefs2003,Nee2005,Ricklefs2006,Allen2007}, one
of the problems with the neutral theory relies on the fact that average
species lifetimes are typically too short, up to some order of
magnitude, compared to those estimated from fossil records. The origin
of such discrepancy is likely due to the point speciation mode,
typically implemented in neutral models \citep{Hubbell2003}. As
suggested by \citet{Allen2007}, a possible way-out for
overcoming this problem is to allow for larger incipient-species
abundances.  Moreover, the same authors have also shown that allowing
for some environmental stochasticity can decrease the lifetime of very
abundant species, which is another issue with the neutral prediction
for the species lifetimes distribution.

Furthermore, our model shows that niche-stabilization induced by
intraspecific interactions larger than interspecific ones, for a
given value of $\nu$, both increases the average lifetime (up to 2 or 3
order of magnitudes depending on $\nu$ and $c$) and suppresses large
fluctuations.  In particular, the smaller the value of $\nu$, the larger is the
effect of stabilization. This allows the system for sustaining
a larger diversity (Fig.~\ref{fig2}) of longer-lived
(Fig.~\ref{fig-mom}) species with respect to a purely neutral
community and, at the same time, suppressing the probability of
species with exceedingly large lifespan with respect to the average.

\section{Acknowledgments}

We thank A. Maritan, M.A. Mu{\~n}oz and S.P. Ellner for comments on
the manuscript.  We thank A. Cavagna for useful suggestions on how to
present the self-consistent derivation.  SP acknowledges partial
support from Spanish research ministry through grant
FIS2012-37655-C02-01.  MC acknowledges support from MIUR
PRIN-2009PYYZM5.

\appendix

\section{Computation of $g$ and $\theta$\label{appendix}}

Here, we derive the expression for the normalization $\theta$
presented in Eq.~(\ref{norm}), and discuss a semi-analytical method to
estimate the parameter $g$, which  is self-consistently
defined as
\begin{equation}\label{gdef}
g=g(c,\nu)=\frac{\widetilde{\Phi}'(0)}{\widetilde{\Phi}'(c)}=
\frac{\sum_{k=1}^\infty (1-\nu)^kg^k e^{-\frac{c}{2}k(k-1)}}
{\sum_{k=1}^\infty (1-\nu)^kg^k e^{-\frac{c}{2}k(k+1)}}\,.
\end{equation}

For the sake of notation simplicity we denote with $N$ and
$D$, respectively, the numerator and denominator in the
right hand side of expression (\ref{gdef}). By rearranging the
indices in the summations, one can show that they satisfy
\begin{equation}\label{numdem}
D=\frac{N-g(1-\nu)}{g(1-\nu)}.
\end{equation}
Substituting (\ref{numdem}) into (\ref{gdef}) yields an expression for
$D$:
\begin{equation}\label{denom}
D=\sum_{k=1}^\infty (1-\nu)^kg^k e^{-ck(k+1)/2}=\frac{1-\nu}{\nu}\,.
\end{equation}

The normalization $\theta$ can be then derived by imposing the
condition $\sum_k k \Phi(k)=J$.  By direct substitution, one obtains
$\sum_k k\Phi(k)=\theta N$. Then, using the fact that
$N=gD=g(1-\nu)/\nu$ yields the result
(\ref{norm}).

Let us now discuss how estimating $g$ in the relevant parameter
range of $c\ll1$ and $\nu\ll1$.  In such limit, one can approximate
very closely the series in (\ref{denom}) with an integral
\begin{eqnarray}\label{g_semi}
&&\frac{1-\nu}{\nu}\approx \int_1^{\infty}dk\ (1-\nu)^kg^k
e^{-\frac{c}{2}k(k+1)}
=\sqrt{\frac{\pi}{2c}} \\
&&\mathrm{erfc}\left(\frac{3c-2\ln(g(1-\nu))}{2\sqrt{2c}}\right)
\exp\left\{{\frac{[c-2\ln((1-\nu)g)]^2}{8c}}\right\}\nonumber
\end{eqnarray}
Equating the first and the last term in the above expression leads to
Eq.~(\ref{g_sem}), that can be solved for $g$ semi-analytically
by standard methods. The dependence of $g$ on $\nu$ and $c$ is 
shown in Fig.~\ref{fig-g}.
\begin{figure}[htb]
\centering
\includegraphics[width=0.4\textwidth]{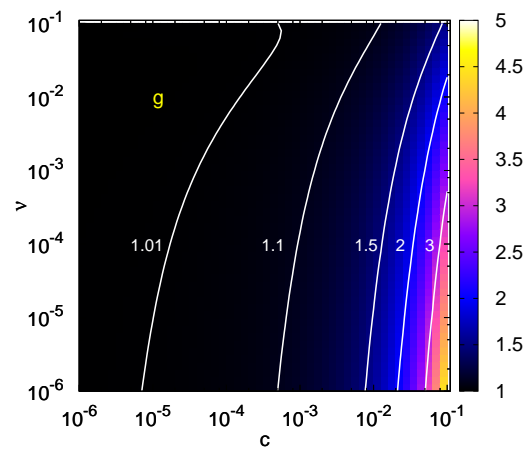}
\caption{Contour plot of the quantity $g$ in the $(c,\nu)$-plane, as computed by solving Eq.~(\ref{g_sem}) semi-analytically.
 \label{fig-g}}
\end{figure}

We conclude by discussing possible analytical expressions for $g$. In
the neutral limit of $c=0$, one has $g=1$, it is thus tempting to
build up an expansion to obtain a closed expression for $g$ at least
in the limit $c\ll 1$. However, such expansion in Eq.~(\ref{g_semi})
requires additional assumptions on the relative magnitude of the two
small parameters $c$ and $\nu$. For example, the argument of the error
function is very small when $\nu\ll c\ll 1$, but very large when
$c\ll\nu\ll1$, so that in the first case a Taylor expansion is
appropriate, while in the second one has to perform an asymptotic
expansion. More formally, one can show that the function $g(\nu,c)$ is
not analytic at the point $\nu=c=0$, so that one cannot perform a
Taylor expansion in the two small parameters. This also suggests that,
more in general, other near-neutral models can be hard (or impossible)
to treat with perturbative methods due to the interplay between the
two small parameters.

\end{document}